\def\beq{\begin{equation}}
\def\eeq{\end{equation}}
\def\bea{\begin{eqnarray}}
\def\eea{\end{eqnarray}}
\def\bq{\begin{quote}}
\def\eq{\end{quote}}

\def\gappeq{\mathrel{\rlap {\raise.5ex\hbox{$>$}}
{\lower.5ex\hbox{$\sim$}}}}

\def\lappeq{\mathrel{\rlap{\raise.5ex\hbox{$<$}}
{\lower.5ex\hbox{$\sim$}}}}

\def\bbz{fa Z \kern-8.9pt Z}

\documentstyle [12pt]{article}
\begin{document}
\thispagestyle{empty}
\vspace{1cm}
\begin{center}
{\large  Cosmological Constraints on B-L Violation } 
\end{center}
\vspace{1cm}
\begin{center}
{Sacha Davidson }\\
\vspace{.3cm}
{CERN Theory Division, CH-1211, Gen\`eve 23, Switzerland}
\end{center}
\hspace{3in}

\begin{abstract}
This is a review of the cosmological bounds on
$B-L$ violating interactions, and the loopholes in the argument
that gives these constraints.  If one assumes that
the baryon asymmetry we observe
today was present above the electroweak phase transition
in equilibrium with the non-perturbative $B+L$
violating processes, then interactions that violate
all three of $\{ B/3 - L_i\}$ cannot simultaneously
be in equilibrium. Otherwise the baryon asymmetry would be washed out.
Therefore violation of at least one
of the  $B/3 - L_i$ must be small. This
argument can be evaded by not having the observed baryon
asymmetry present in the thermal bath (for instance,
make it at the electroweak phase transition), by
using a non-standard cosmological model, or
possibly by some mass effects in models
where the difference between lepton flavour
asymmetries is conserved.
\end{abstract}

\section{introduction}

It is observed that at least up to the scale
of our local galaxy cluster, the Universe we
see is made of matter rather than anti-matter.
It is difficult to build a cosmological
model where the Universe contains  equal numbers of
baryons and anti-baryons  separated on
scales larger than galaxy clusters
(There is a problem with causality \cite{causality}),
so one usually assumes that
the Universe as a whole contains a net
excess of baryons over anti-baryons,
and that at some time in the early history
of the Universe, this asymmetry was generated.
This is the Baryon Asymmetry of the Universe,
or BAU (for reviews, see, $e.g.$  \cite{BAUrev,RS,KT}.)

There are three ingredients required to
create a baryon asymmetry \cite{Sakharov}:
baryon number violation, C and CP violation,
and some out of equilibrium dynamics.
The Standard Model (SM) does contain
all these ingredients, but not in 
the right quantities to generate
the BAU.

The Standard Model violates C maximally
(because it is chiral), and  contains small
amounts of CP violation in the CKM matrix
(probably not enough to generate the
BAU though).
It has no perturbative baryon number
violation. The lowest dimensional
baryon number violating operator allowed by
standard model gauge symmetries and
involving SM fields  is
$qqq \ell$ \cite{Ross}, which is of dimension six.
However the Standard Model does contain
non-perturbative baryon number violation \cite{'tHooft}: 
the  current associated with the quantum number
$B+L$ (baryon + lepton number) is anomalous,
so certain gauge field configurations
can induce the  $B+L$ violating
vertex $(qqq \ell)^3$ (see, for instance,
\cite{RS} for a review and references).
This operator allows processes such as
\beq
u_g  d_b  d_r 
c_g s_b s_r
 t_g  b_b b_r \longrightarrow 
\bar{\nu}_e \bar{\nu}_{\mu} \bar{\nu}_{\tau}
\eeq
where $g,r,b$ are SU(3) colour labels, not summed, and the quarks
are SU(2) doublet members. 
 At zero temperature
the rate for such $B+L$ non-conserving processes
is exponentially suppressed by the action of
the weak gauge field configuration,
 so one does not
expect to see low energy $B+L$ violation.
At finite temperature 
the  rate is much less suppressed;
the rate above the electroweak phase transition
is
\beq
\Gamma_{B+L   \! \! \! \! \! \! \! / }\,\,\, = \kappa \alpha^{n} T
\eeq
where $\kappa$ is a constant of order 1, and $n$
is 4 or 5 \cite{Arnold}. This is clearly
in thermal equilibrium  in the early
Universe ($\Gamma >H \equiv$ the expansion
rate of the Universe; if $\Gamma >H $ the
timescale for the interaction is less than
the age of the Universe.).
After the phase
transition, the rate is Boltzmann suppressed
($\Gamma \sim \exp \{ -m_W/(\alpha T) \} $),
so rapidly drops out of equilbrium.
This is potentially very interesting for
baryogenesis (as was noted by \cite{KRSa}): 
a source of baryon number violation
that is efficient in the early Universe when
the asymmetry needs to be generated, and then turns
off at low energy in perturbative processes
where we see no baryon number violation.

Unfortunately the Standard Model lacks
the third ingredient required to generate
the baryon asymmetry.
There is not a big enough departure
from equilibrium  at the electroweak phase
transition; the $B+L$ violating processes turn off
gradually and no  asymmetry in generated \cite{EPT}. 
The observation of the baryon asymmetry
can therefore be interpreted as evidence for new
physics, since it does not appear possible
to generate an asymmetry within the
Standard Model.  

The new physics can have two effects on the generation 
of the baryon asymmetry.
It may simply add enough CP violation and
make the phase transition sufficiently strongly first order
that the baryon asymmetry is generated at the
phase transition using the SM $B+L$
violation. Alternatively, 
the beyond-the-Standard Model physics may involve new
sources of $B$ and/or $L$ violation 
which generate the observed asymmetry
before the electroweak phase transition.
In the first case there are no cosmological
bounds on $B-L$ violation. So for most
of this review, I will assume the second 
possibility: suppose that the
baryon asymmetry we observe today was generated
early in the history of the Universe and
is present as an asymmetry in the number density of
quarks and antiquarks in the thermal bath just
above the electroweak phase transition. 
The anomalous $B+L$ violating processes are also
operating in this plasma, so if the asymmetry generated
previously wishes to survive in their presence, 
it must carry at least one of 
$B/3 - L_e$, $B/3 - L_{\mu}$ or $B/3 - L_{\tau}$. These
are the three global quantum numbers exactly conserved
in the Standard Model. If there are simultaneously interactions in 
equilibrium  that violate all three of
these  quantum numbers, all asymmetries will be washed out.
Since we are assuming that the baryon
asymmetry we observe today was generated
before the electroweak phase transition,
this is undesirable. One can avoid this outcome
by requiring that processes violating
one of the $ \{ B/3 - L_i \} $ be out of
equilibrium:
\beq
\Gamma_{B/3 -L_i} < H ~~~~~{\rm for~one}~i \label{roughbound}
\eeq
where $H$ is the expansion rate of
the Universe, and $i$ is one of $e, \mu$ or
$\tau$.  This condition should hold for
all temperatures during which the baryon asymmetry
is at risk: 100 GeV $ \lappeq T \lappeq  100 $ TeV.
The reasons for this range of temperatures
will be discussed in section 3.

 For a  
yukawa-type $B-L$ violating
interaction $\lambda \phi \psi \psi$
(eg R-parity violating trilinears
in Supersymmetry),
equation (\ref{roughbound})  becomes 
\beq
10^{-2} \lambda^2 T < \frac{25 T^2}{m_{pl}}
\label{yuk}
\eeq
which gives a bound of order
$\lambda < 10^{-7}$. I use $H = 10 T^2/m_{pl}$
in the SM, and
$H = 25 T^2/m_{pl}$ for SUSY particle
content.
One can estimate the rate associated with
a majorana neutrino mass to be of order
\beq
\Gamma_{m_{\nu}}  \sim \frac{ 10^{-3} m^2}{T} < H
\label{numass}
\eeq
which gives 
  $m_{\nu} < 30 $ keV
in one generation. The numerical
factors in these bounds will be discussed more
carefully in section 2.  Bounds can also be 
set on non-renormalisable 
$B/3 - L_i$ violating operators of
dimension $D = 4+n  > 4$; if all the coupling
constants  are absorbed
into the mass scale $M$, the rate can be
estimated as
\beq
\Gamma \sim 10^{-3} \frac{T^{2n +1}}{M^{2n}} < \frac{10 T^2}{m_{pl}}
\label{nonren}
\eeq
For $D = n+4 = 6$, this gives
$M > 10^4 T_{{\rm GeV}}^{3/4}$ GeV. It is clear that non-renormalisable
interaction rates increase faster with
temperature than the expansion rate, so
are more likely to be in equilibrium
 at higher temperatures. The maximum
temperature  at which the bounds can be applied,
which gives the best bounds
($T \sim 100$ TeV), will be discussed in section 3.

These constraints are comparatively stringent,
so it us useful to know exactly where they apply and how
they can be avoided. These bounds apply to only
one of the $ \{ B/3 - L_i \}$, so lepton number violation 
in one family and baryon number
violation must be small.  In section 2
I will review estimates of interaction
rates for various coupling constants,
and the bounds one can derive from these
estimates.  The most obvious loophole is to generate
the asymmetry at the electroweak phase
transition, or more precisely, to not
have the baryon asymmetry present in the plasma
 just above the EPT. In
section 3 I will show that the BAU
only needs to be absent from the thermal
soup for temperatures between $\sim 100$ TeV
and the EPT,
and discuss various  ways of 
accomplishing this. 
 Section 4 contains various other mechanisms for 
avoiding the bounds such as low $T_{reheat}$
inflationary models. There is a summary of the
constraints and loopholes in section 5. 
The remaining part of the introduction is a short
overview of previous work on this topic. 

It has been known for a long time that if the baryon
asymmetry is generated by the out-of-equilibrium
decay of heavy GUT particles, then baryon number
violating processes mediated by these
particles must be out of equilibrium after
the asymmetry is generated \cite{GUTBAUrev}.
Otherwise they will wipe out the asymmetry
generated in the decay. 
For the baryon asymmetry to survive
in the presence of the $B+L$ violating
electroweak effects,  it must be an asymmetry in 
$B-L$, which makes it difficult (but not impossible, see 
section 4.1) to preserve
the baryon asymmetry in $B-L$ conserving GUTs.
An attractive alternative, introduced by
Fukugita and Yanagida\cite{FYBAU}, is to generate a 
lepton asymmetry in the decay of heavy
singlet neutrinos, and use the electroweak $B+L$ violation
to transform this into a baryon asymmetry. 
Lepton number violating interactions
mediated by the heavy singlet neutrinos
must be out of equilibrium after the lepton
asymmetry is generated, to avoid washing
out the asymmetry\cite{FY}.  This bound on
lepton number violating processes is
not limited to leptogenesis scenarios;
 Barr and Nelson \cite{NB}, and \cite{FGLP},
 observed that the baryon asymmetry,
the $B+L$ violating processes, and interactions
that violate the three $\{ B/3 - L_i \}$
cannot simultaneously coexist in equilibrium, and used
this to set bounds on majorana neutrino
masses in one generation.  This argument
was then extended to R parity violating
interactions in the Supersymmetric version
of the Standard Model \cite{FGLP,CDEO1}
and to generic $B-L$ nonconserving interactions
\cite{CDEO1}. The constraints and loopholes
were then discussed by many people
\cite{CDEO2,DR,CDEO3,IIMS,other}.

\section{rate estimates and  constraints}

The argument that gives cosmological bounds
of $B-L$ violation goes as follows: assume that the
observed baryon asymmetry was generated before the EPT.
To survive in the thermal soup, it must 
carry a quantum number that is ``effectively 
conserved''---{\it i.e.} conserved by
the interactions that are in 
chemical equilibrium. 
The SM interactions are all in chemical
equilibrium just above the EPT, and the three
global quantum numbers that they conserve
 are the $\{ B/3 - L_i \}$.
To protect the BAU, any additional interactions present
must also ``effectively conserve'' at least one 
of the $\{ B/3 - L_i \}$, or equivalently,
interactions violating at least one of
the  $\{ B/3 - L_i \}$ must be out of equilibrium. 

In the first part of this section,
I will estimate interaction rates
associated with coupling constants
of different mass dimensions. It
follows from these estimates that all the
SM interactions come into equilibrium
before the EPT. In the
second part, I will discuss the equations
of chemical equilibrium 
and charge conservation for the
SM interactions. These are
interesting because there are
small ``second order''  terms
proportional to lepton masses
that can be used to protect an asymmetry.
Finally in the third subsection
I will use the rate estimates
from section 2.1  to set bounds on
neutrino masses, $R$-parity
violating coupling constants, etc.

 As one
can see from the estimated bounds
in equations (\ref{yuk}), (\ref{numass}) and  (\ref{nonren}),
the best constraints are on
renormalisable couplings.
With the SM particle
content, there are no $B-L$ non-conserving
operators of dimension less than 5. 
However, in the supersymmetric extension
of the SM, when $R$-parity is not imposed,
baryon and lepton number violating masses and
trilinear terms are possible. In this proceedings,
I will discuss constraints on $B-L$ violation
with SM particle content at the electroweak
scale, and also with SUSY particle content.

The MSSM superpotential is
\beq
W = \mu H_1 H_2 + h_u^{pq} H_2 Q_p U^c_q + h_d^{pq} H_1 Q_pD^c_q +
    h_e^{ij} H_1 L_i E^c_j ~~~. \label{W}
\eeq
The Lagrangian also contains kinetic terms,
gauge interactions, D-terms and soft SUSY breaking
terms of the form
\beq
{\rm soft~ masses} + B_H H_1 H_2 +  A_u^{pq} H_2 Q_p U^c_q + 
 A_d^{pq} H_1 Q_pD^c_q +
    A_e^{ij} H_1 L_i E^c_j ~~~. \label{soft}
\eeq
I am abusively using capital letters for both
superfields (as in eqn \ref{W}) and scalar
component fields (as in eqn \ref{soft}). Quark
generation indices are $p,q,r,s...$ and
lepton indices are $i,j,k...$. Whether indices
are up or down makes no difference. 

The above SUSY Lagrangian is with the symmetry $R$-parity
imposed. Other 
renormalisable interactions 
can be constructed with the
MSSM particle content, but these
new interactions violate either $B$ or $L$,
and are often removed by imposing a symmetry like 
$R$-parity \cite{herbi}. See \cite{IR} for
other possibilties.  These are examples of the
interactions I am interested in setting bounds
on. There are possible new superpotential terms
\beq
W_{R  \! \! \! \! /} = \epsilon^i H_2 L_i + \lambda^{ijk} L_i L_j E^c_k
                + \lambda^{'ipq} L_i Q_p D^c_q 
                  + \lambda^{''pqr}  U^c_p D^c_q D^c_r \label{WR}
\eeq
and new $B$ or $L$ violating soft terms
\beq
\begin{array}{l}
{\rm soft~ masses~ mixing}~ L^{\dagger}~ {\rm and}~ H_1 \nonumber \\+  
 B^i H_2 L_i  + A^{ijk} L_i L_j E^c_k
                + A^{'ipq} L_i Q_p D^c_q 
                + A^{''pqr}  U^c_p D^c_q D^c_r      
                ~~~. \\
\end{array}
\label{Rsoft}
\eeq

I assume that SUSY masses are of order 100 GeV $\sim T_c$.
The bounds  on trilinears
are not substantially weakened by larger soft
masses---if one sets the constraints at $T \sim m_{SUSY} 
\sim 1$ TeV, they are weakened by a factor of about 3. 
The effects of larger SUSY masses were discussed 
in \cite{DR,IIMS}.

\subsection{rate estimates}

In this section I will roughly estimate
interaction rates for a selection of operators.
See for instance 
\cite{CDEO2,DR,IIMS} for  more thorough discussions
that disagree in details but all get 
approximately the same constraints.
For a detailed calculation of the
interaction rate associated with a 
specific coupling constant see, {\it e.g.}
\cite{CKO}.

The rate for a  boson to 
scatter off a fermion 
via the yukawa interaction $ y \phi \bar{\psi}_1 \psi_2$
can be estimated using kinetic theory
(see {\it e.g.} 
\cite{KT} for an introduction) to be \cite{CDEO2}
\beq
\Gamma(\phi \psi_1 \rightarrow  \phi \psi_1) 
\simeq < \sigma v n_{\psi_1}> \sim  4\times 10^{-4}y^4 T \label{gb}
\eeq
where $\sigma$ is the $T=0$ cross section,
$n_{\psi_1}$ is the equilibrium thermal distribution
of fermions $\psi_1$,  $v$ is
the relative velocity of the
 boson and the fermion $(\simeq c = 1)$,
and the brackets represent
averaging the cross section over
thermal distributions in momentum space.

For a gauge boson, the matrix element, and therefore
the numerical coefficient, will be different,
but if one substitutes the gauge coupling $g$ 
for the yukawa $h$ in
equation (\ref{gb}), it  is clear that gauge
interactions are in equilibrium
up to temperatures of order
$\lappeq m_{GUT}$.

One can similarly estimate the rate for
$\phi \psi_1 \rightarrow \psi_1 \gamma$ where $\gamma$
is an arbitrary gauge boson, to be of order
\beq
\Gamma(\phi \psi_1 \rightarrow  \gamma \psi_1) 
 \sim  5 \times 10^{-3}y^2 \alpha T ~~~~.\label{gb2}
\eeq

The rate for a scalar (Higgs or spartner) to
decay to two fermions via the interaction
$ y \phi \bar{\psi} \psi$ can be estimated as
\cite{CDEO2} 
\beq
 \Gamma \sim  1.4 \times 10^{-2} \frac{y^2 m_{\phi}^2}{T}~~~~.
\label{scal}
\eeq
$m_{\phi}$ is the decaying particles mass
in the finite temperature effective potential. 
Equation (\ref{scal}) neglects final state
phases space suppresion due to fermion masses, which can be
significant if the mass of the scalar is of order $gT$.
(The final state fermions can also
 have thermal masses of order $gT$.)

To estimate a rate associated with a yukawa,
one therefore has a choice between
scattering and decay.
The scattering rate has an extra power of $\alpha$,
but the decay may be suppressed by the small
amount of final state phase space,
depending on the (thermal) masses of
the participating particles.
I will use the decay rate estimate to set
bounds on $B-L$ violating couplings,
because the decaying particle is a
SUSY spartner with a zero temperature mass
of order 100 GeV.

In the SM case, $y \simeq  \sqrt{2} m_f/v$
where $v = $ 246 GeV is the Higgs vev, and
$m_f$ is a fermion mass. (This is not entirely
accurate for the quarks, because the finite
temperature mass eigenstates will not be
the same as the zero temperature ones,
but this should  be a reasonable estimate).
Taking $H \sim 10 T^2/m_{pl}$, one finds
that the electron yukawa is in thermal
equilibrium below $T_{e_R} \sim 10-100$ TeV \cite{CDEO3},
 the $u_R$ quark will
be in chemical equilibrium below
$T_{u_R} \sim 100$ TeV, and so on.
(See \cite{CKO}
for a careful discussion of
when the electron yukawa comes into
equilibrium,  taking  into account
the thermal masses in the decay rate.).
This means that all SM interactions
are in equilibrium just above the EPT.

To estimate a  rate associated with a majorana
neutrino mass, I need to consider how
it is generated. If the zero temperature
mass  arises from the dimension 5 operator
$y^2 (\ell \ell H H)/M$, where $M$ is a heavy 
right-handed singlet neutrino mass,  then
at temperatures $ T_c < T \ll M$
this operator generates two-fermion-two-Higgs
scattering. The rate can be estimated \cite{NB,FGLP}
as in equation (\ref{gb}) to be
\beq
\Gamma(\ell H \rightarrow \ell H)
 \simeq  4 \times 10^{-4} \frac{y^4 T^3}{M^2} ~~~.
\label{mD2M}
\eeq
This is a conservative estimate of the
lepton number violating rate, because
the latter should also include the processes
$\ell \ell \rightarrow HH$, 
$HH \rightarrow \ell \ell $ and so on.

Alternatively, if the majorana mass is
present above the EPT as a mass, then  it
generates lepton number violating processes
such as $\nu W \rightarrow \bar{\nu} W$.
Majorana mass insertions on the external
neutrino legs contribute the lepton number
violation. The mass corrections to
the scattering process  can be estimated from (\ref{gb}) as
\beq
\Gamma(\nu W \rightarrow \bar{\nu} W)
 \simeq   10^{-3} g^4 T\left( \frac{ m_{\nu}}{T} \right)^2 
\label{mnu}
\eeq
and to a decay as
\beq
\Gamma( W \rightarrow {\nu} \nu) \sim
   3 \times 10^{-2} g^2 T  \left( \frac{ m_{\nu}^2}{T^2} \right) 
\label{mnu2}
\eeq
(To make this estimate, I took $m_W^2(T) 
\sim g^2T^2/3$ and $ m_{\nu}^2(T) \sim g^2T^2/8$.
For these masses, the decay $ W \rightarrow {\nu} \nu$
is kinematically not allowed, so I ought not
to use this estimate. It is nonetheless in the literature. )

\subsection{chemical equilibrium equations}

There are two approaches  in the literature
to the equations of chemical
equilibrium  as applied to cosmological
contraints on $B-L$ violation. The more
common one uses  kinetic theory
\cite{HT,DR}, and is reliable
insofar as it agrees with the more
esotheric calculations that start from
the free energy \cite{KS}. I will
use something in between,
that I think gives the right answers
 as long as one is not to near the EPT
(See \cite{KS} for an analysis that
is valid for all temperatures.)

If an interaction is in chemical equilibrium,
for example the decay 
$\phi \rightarrow \psi_1 \psi_2$, then
the sum of the chemical potentials of
the participating particles is zero:
\beq
\mu_{\phi} - \mu_{\psi_1} - \mu_{\psi_2} = 0 ~~~~.
\eeq

Above $T_c$, all the
gauge
bosons have  zero chemical potential
\cite{HT,DR,DKO}. From equation (\ref{gb})
it is clear that SM gauge interactions
are in equiliibrum up to
very high temperatures, so all members of a gauge
multiplet have the same chemical
potential.

At $T \simeq T_c$, all the SM processes are
in  equilibrium. If the chemical potential
for a gauge multiplet is written as
the name of the multiplet (so the chemical
potential for $e_R^i$ is $e_R^i$), then the
Yukawas imply
\beq
\begin{array} {ll}
-e_R^i  + \ell^i + H & = 0 ~~~, \\
-u_R^i  + q^j - H & = 0  ~~~,\\
-d_R^i  + q^j + H & = 0  ~~~. \\
\end{array} 
\label{Higgs}
\eeq
The rate estimates in section 2.1 say that
the electron yukawa is in equilibrium 
below $T_{e_R} \sim 10-100$ TeV. So
equations (\ref{Higgs}) apply for $T_c < T < 10-100$ TeV.
Above this temperature, the $e_R$ have a thermal distribution (they
have gauge interactions) but any asymmetry in the $e_R$
is decoupled from the rest of the plasma. In the
absence of an interaction that can transfer the $e_R$
saymmetry to other particles, asymmetries must remain
amoung the other particles (in particular, a baryon asymmetry)
to cancel the hypercharge carried by the
$e_R$ asymmetry.

The non-perturbative $B+L$ violating processes
give
\beq
9 q_L + \sum_i \ell^i = 0
\label{B+L}
\eeq
Note that these interactions eat three units of
baryon number and three units of lepton number,
so they violate $B+L$ but not $B-L$.  However,
when they are in equilibrium with SM interactions
in the early Universe, they imply
$n_B = - \frac{28}{51} n_L
$ and not $n_B =- n_L$. This is because the
interactions in equilibrium try to minimise
the free energy, and the minimum is not
neccessarily where all the quantum numbers are zero.
The ``sphalerons'' of equation (\ref{B+L}) do not
take $n_{B+L}$ to zero because they
only eat SU(2) doublets, and the SU(2) singlets
(who also carry $B$ or $L$) have different hypercharge.
In a plasma with $Y = 0$ but non-zero
$n_{B-L}$, the free energy minimum is at
$n_B = -(28/51) n_L$ rather than at $n_{B+L} = 0$.

In the case of the MSSM, there are many new
particles. The gauginos have majorana masses (sufficiently
large to be in chemical equilibrium) so can
carry no asymmetry and
must have zero chemical potential. 
This means that the sfermion and
fermion chemical potentials are
equal and opposite to each other,
because the  gaugino interaction
$\tilde{g} \phi \psi$ imply
\beq
\phi + \psi + \tilde{g} = \phi + \psi = 0
\eeq
The two Higgs have opposite chemical potentials
due to the mass term $\mu H_1 H_2$,
so one gets the same
chemical potentials as in the SM, subject
to the same constraints due to yukawa couplings. 

If the flavour off-diagonal slepton masses are large enough for lepton
flavour violation to be in equilibrium,
then only $B-L$ is conserved, and not
the three $B/3 -L_i$ separately. I will
assume  for the moment that the off-diagonal
masses are ``small enough''
($m^2_{ij}/m^2_{ii} < 10^{-2} $), and come
back to this constraint when I am
discussing bounds on $B-L$ violating
operators in supersymmetric theories.

This gives six equations for the 10
unknown chemical potentials. The four
free parameters correspond  to the
four conserved charges:
 electric charge (or hypercharge)
and the three $\{ B/3 - L_i\}$. 
The electric charge must be zero
(imposing $Y=0$ gives an equivalent constraint):
\beq
\sum_{particles} Q_{em} (n - \bar{n}) = 0
\eeq
where ($\bar{n}$) $n$ is the (anti-) particle
number density. $n - \bar{n} = \partial F/\partial \mu$
where $F$ is the free energy  and $\mu$ the particle chemical
potential. One finds
\beq
 n - \bar{n} = \frac{2 g}{\pi^2}\mu T^2 \left\{
\begin{array}{ccc}
 1 + { O}( m^2/T^2) & ~ & {\rm fermions} \\ 
 2 + { O}( m^2/T^2) & ~ & {\rm bosons} \\ 
\end{array}
\right.
\label{nbarn}
\eeq
for $m \ll T$.  $g$ here is the
 number of degrees of freedom
of the field---2 for a chiral
fermion and a charged scalar.
One can correctly
get the first term  of equation (\ref{nbarn})  
using  kinetic theory,
by expanding the equilibrium particle
distributions in $\mu$ and $m^2$ (see section 4). However,
this expansion does not always give
the right coefficient for the ${ O}(\mu m^2)$ terms.
Note that I am working above
the EPT, so the masses in (\ref{nbarn})
are thermal masses or soft SUSY masses,
but are not due to the Higgs vev. 
  Neglecting for the moment
the mass effects, one finds in the SM
(with $n_H$ Higgs doublets)
\beq
Q_{em} \propto 3 q_L + 6 u_R - 3 d_R - \sum_i [\ell^i + e_R^i] - 2 n_H H
    = 0
\label{Qem}
\eeq

In the MSSM, the numerical coefficients
change slightly, assuming $m_{SUSY} \lappeq T_c$,
because there are now two Higges and spartners
as well as partners in the soup. The net effect
is to change the number of Higges $n_H$ from one
to two in (\ref{Qem}).

There are now seven homogeneous equations
for the 10 unknown chemical potentials.
If one fixes the asymetries in the three
$\{ B/3 - L_i \}$s, all the chemical
potentials are determined. 

The baryon asymmetry carried by the quarks is
\beq
\begin{array}{ll}
n_B  & =  \sum_{col} \sum_{gen} (2 \times \frac{1}{3}
(n_{q_L^i} - n_{\bar{q}_L^i}) +\frac{1}{3}
(n_{u_R^i} - n_{\bar{u}_R^i})  +\frac{1}{3}
(n_{d_R^i} - n_{\bar{d}_R^i})) \\
& \propto 12 q_L T^2  \\ 
\end{array}
\label{1}
\eeq
where I have used equations (\ref{Higgs})
and neglected $O(\mu m^2)$ corrections.
Quarks of different families have the same
chemical potential so quark mass effects in equation
(\ref{nbarn}) cannot be used to preserve an asymmetry. 
Equation (\ref{1}) can be written using equations
(\ref{Higgs}), (\ref{B+L}) and (\ref{Qem}) as 
\beq 
n_B = \frac{24 + 4 n_H}{66 + 13 n_H} \sum_i n_{B/3-L_i} +  
\frac{47 y_{\tau}^2}{1896 \pi^2}(n_{L_e - L_{\tau}} +  n_{L_{\mu} - L_{\tau}})
\label{masseffects}
\eeq
where $n_H$ is the number of Higgs doublets. 
I have included here the lepton mass effects---specifically
the $\tau$ yukawa $y_{\tau} = m_{\tau}/(175~ {\rm GeV})$, 
because it is largest. These small
corrections, which are present
below \cite{KRS} and above\cite{DKO} the
EPT, will be discussed in section four. It is
clear that if $B-L = 0$, one needs
very large lepton flavour asymmetries 
($n_{L_i - L_{j}}/s \sim 10^{-3}$,
where $s$ is the entropy density)
to preserve the baryon asymmetry via
lepton mass differences. (As noted in
\cite{DR}, slepton mass differences
could be more effective.)

\subsection{bounds on $B-L$ violating coupling constants}

Equation (\ref{masseffects}) says, as
expected, that a baryon asymmetry will remain in
the plasma if an asymmetry remains in any one
of the $\{ B/3 - L_i \}$. Neglecting
the lepton mass effects, it is clear that
we need an asymmetry in at least one of the
$\{ B/3 - L_i\}$ to preserve the asymmetry
in the quarks. This means that interactions violating
one of the $\{ B/3 - L_i \}$ need
to be out of equilibrium.  Suppose,
for instance that it is $B/3 - L_e$ that
is not washed out. Then one
requires
\beq
\Gamma_{B/3 - L_e} < H 
\label{bound}
\eeq
for $T_c < T < T_{e_R} \sim 100$ TeV.

I can translate (\ref{bound})
into a constraint on coupling constants
using the rate estimates from section 2.1.

For majorana neutrino masses
 that are present in the thermal soup above the EPT as a mass,
requiring $\Gamma$ from equations
(\ref{mnu2}) or (\ref{mnu}) to be less than $ H$ at $T = 100$ GeV gives
\beq
 m_{\nu} < 10 - 100~ {\rm keV}
\eeq
(The weaker bound corresponds to the 
scattering interaction rate).

In the case where $m_{\nu} \sim m_D^2/M$,
above the EPT there is a lepton number violating
scattering process mediated by the
dimension 5 operator $(\ell \ell HH)/M$.
As
discussed in the introduction,
  $\Gamma_{L \! \! \! \! /}\,\,\,/H$
increases with the temperature for $D>4$
operators, so the best bound is at the highest possible
temperature. This is $T_{e_R} 
\sim 100$ TeV \cite{CDEO3}. 
Requiring  equation
(\ref{mD2M}) to be out of equilibrium at
$T \sim 100$ TeV gives the bound
\beq
m_{\nu} \lappeq {\rm keV} ~~~~.
\eeq

These  are bounds on a neutrino mass
that violates electron lepton number.
So in the basis where the charged lepton
mass matrix is diagonal the majorana
neutrino mass matrix entries $m^{e i}_{\nu} < 10 $
keV for $i = e, \mu, \tau$.

The $R$-parity violating
masses $\epsilon_i \tilde{h}_2 \ell_i$
are slightly trickier \cite{DE1,DE2},
because one can set them to zero
by redefining the higgsino $\tilde{h}_1$.  
The ``problem'' here is that the
${H}_1$ superfield has the same
gauge quantum numbers as
 the lepton superfields ${L}_i$.
In a lepton number conserving theory,
${L}_i$ are distinct from
${H}_1$ because they carry
lepton number. However if the $i$th
lepton number
is not conserved, then there
is no unique definition of which linear
combination of ${L}_i$ and ${H}_1$
is ``the Higgs''.

For setting bounds on lepton number violation,
it should not matter which basis we calculate in.
Whether an asymmetry in, for instance $L_e$,
survives in the plasma is a physical
question that should not be basis dependent.
It is possible to construct 
basis independent ``invariants''
that parametrise the amount of
$R$ parity violation between
different coupling constants
(analogous to Jarlskog invariants
for $CP$ violation), and to
set bounds on the invariants
\cite{DE1,DE2,D}. However, one
can get the right answer with
a less formal approach.

If we rotate the superpotential term
$\epsilon_i {H}_2 {L}_i$ 
into $\mu {H}_1 {H}_2$
by redefining ${H}_1$:
\beq
{H}_1' = \frac{1}{ \sqrt{ \mu^2 + \epsilon^2} } 
(\mu {H}_1 + \epsilon_i {L}_i)
\label{BBB}
\eeq
then we generate new trilinears:
\beq
\frac{h_d^{pq} \epsilon_i }{\sqrt{\mu^2 + \epsilon^2}} L_i Q_pD_q^c
~~~~~~~
\frac{h_e^{jk} \epsilon_i }{\sqrt{\mu^2 + \epsilon^2}} L_i L_jE_k^c
\eeq

If the soft masses $B_H H_1 H_2 + B_i H_2 L_i$
are parrallel (in $H_1, L_i$ space) with $(\mu, \epsilon_i)$,
({\it i.e.}  $B_i/B_H = \epsilon_i/\mu$),
then they are simultaneously rotated away. If
in this basis the soft masses mixing  $H_1$
and the $L_i$ are also zero, then 
 there is no lepton number violation in the masses.
In this case, there are two possible
bounds one can set on $\epsilon_i$: one could
require that the rate associated with the mass
$\epsilon_i$ be out of equilibrium,
or that the trilinear  interaction
$(h_d^{33} \epsilon_i /{\sqrt{\mu^2 + \epsilon^2}})^2 T < H.$
The first bound would imply $\epsilon_i < 10$ keV,
the second $\epsilon_i < 10^{-5} \mu \simeq $ MeV. 
So it matters which is the right one to impose. 
It turns out that the second, weaker constraint
is the correct one. One can
see this in a one generation model;
the stronger rate due to the mass term
chooses the Higgs to be the direction of
equation (\ref{BBB}), so the lepton is
the orthogonal direction,
and lepton number is violated by
the $\lambda' LQD^c$ term
in this basis. One can check that in the
basis (\ref{BBB}), the trilinear $\lambda^{'i33}$
can be written in terms of the coupling constants
of the original basis as 
$(\epsilon_i h_d^{33}- \mu \lambda^{'i33})/ {\sqrt{\mu^2 + \epsilon^2}}$.
Equation (\ref{CCC}) gives the bound
\beq
(\epsilon_i h_d^{33}- \mu \lambda^{'i33})/ {\sqrt{\mu^2 + \epsilon^2}} 
< 10^{-7} ~~~{\rm for~one~} i 
\label{DDD}
\eeq
or $\epsilon_i < $ MeV for  $\mu \sim 100$ GeV.

It is unlikely in realistic models that
$(\mu, \epsilon_i)$ should be exactly
parrallel to $(B_H, B_i)$, or that
the soft masses $m^2_{H_1 L_i}$ should
be zero in this basis. 
Small lepton number violating masses
parametrising the misalignment
should remain.   For simplicity, let
me neglect the $m^2_{H_1 L_i}$, and
just consider $\epsilon_i$ and $B_i$.
The rate for these masses can be estimated as
a mass correction  times a gauge interaction
rate, like  the estimate for the neutrino majorana mass,
that is
\beq
\Gamma \sim {\rm min} \left\{ \frac{ \epsilon_i^2}{\mu^2 + g^2 T^2} ,
~ \frac{ (B_i)^2}{(B_H + g^2T^2)^2} \right\} \times 10^{-2} g^2 T
\label{26}
\eeq
where in this formula $\epsilon_i$ ($B_i$) is taken in the
basis where $B_i$ ($\epsilon_i$) is zero.
The denominator contains a crude attempt
to include thermal mass effects.
To see why equation (\ref{26}) is the right bound, imagine
sitting in the early Universe above the EPT. As the
temperature drops, more and more renormalisable 
interactions come into equilibrium. At some point the
larger of the two mass terms $\sqrt{(B_H)^2 +  (B_i)^2}$
and $\mu^2 +  \epsilon_i^2$ comes into equilibrium, 
and thereby chooses the Higgs direction---{\it i.e.}
if $\sqrt{(B_H)^2 +  (B_i)^2} > \mu^2 +  \epsilon_i^2$
 the Higgs will be combination $B_H H +  B_i L_i$.
then in this basis chosen by the interactions,
one needs the lepton number violating rate associated
with the mass
$\epsilon_i$ to be out of equilibrium,
which is equation (\ref{26}).

 The bound (\ref{26}) can be expressed in a
more basis independent way as:
\beq
\frac{B_H \epsilon_i - \mu B_i}
{\sqrt{(B_H)^2 + (B_i)^2} \sqrt{\mu^2 + \epsilon_i^2}} 
< 2 \times 10^{-7} ~~~,
\label{Bmu}
\eeq
or, in the basis where $B \epsilon_i$ is zero,
$\epsilon_i/\mu < 2 \times 10^{-7}$.
So the cosmological bounds require
that $(\mu, \epsilon_i)$ be more aligned with
 $(B_H, B_i)$ than with the yukawas.

In the supersymmetric Standard Model, lepton
flavours are unlikely to be separately conserved.
The slepton mass matrix is  unlikely to be
exactly diagonal in the lepton mass
eigenstate basis. If lepton flavour violating
processes due to flavour off-diagonal slepton masses
are in thermal equilibrium, then all $B-L$
violating processes need to be out of equilibrium,
not just interactions violating on of the
$B/3 - L_i$, {\it i.e.}  the bounds
(\ref{DDD}), (\ref{CCC}) and (\ref{Bmu})  would apply to
all $B-L$ violating coupling constants, 
and not just those involving one lepton generation.
So if one wants to use the ``flavour loophole'',
and only impose the bounds on one generation,
the flavour changing soft masses must be out of
equilibrium.

The rate associated with a flavour  off-diagonal
soft mass can be estimated  analogously 
to the  previous discussion about $\epsilon_i H_2 L_i$.
There are three interactions that choose
a basis in lepton flavour space: the lepton yukawa $h_e^{ij}$,
the $A$ term $A_e^{ij}$, and the soft masses
$m^2_{ij}$. If I want to separately conserve
lepton flavours in the thermal soup, then all
the interactions in equilibrium must agree
on the basis in lepton flavour space. Suppose
I start in the basis where  $h_e^{ij}$
is diagonal, because this is a familiar basis
from phenomenology, and I put a hat
on the coupling constants in this basis. There will be off-diagonal
terms $[\hat{m}^2]_{ij}$,$\hat{A}_e^{ij}, i \neq j$  in this basis.
Since the interaction rate 
associated with $m^2$ is larger
at $T \sim 100$ GeV than  the rate for $h_e$,  this
is probably not a sensible basis.  I should
rotate to the basis where $m^2$ is diagonal,
because the stronger interaction will choose
the lepton flavour basis. In this new
basis, there will be flavour off-diagonal
yukawas $h_e^{ij} \sim \hat{h}_e^{ii} \hat{m}^2_{ij}/\hat{m}^2_{ii}$
($i \neq j$, and no sums on repeated indices).
These yukawas need to be out of equilibrium:
\beq
\Gamma \sim 10^{-2} \left( \hat{h}_e^{ii} 
\frac{\hat{m}^2_{ij}}{\hat{m}^2_{ij}} \right)^2 T < H
\eeq
If one wants to preserve $B/3 - L_e$, this
gives the bound $\hat{m}^2_{ej}/\hat{m}^2_{ee} < 5 \times 10^{-2}$.
Preserving an different lepton flavour 
would give a stronger bound (the yukawa
is larger) on different elements of $\hat{m}^2$.

Now consider the $A$ term.  Suppose I estimate
the associated rate from equation (\ref{scal}) to be
\beq
\Gamma_A \sim 10^{-2} \frac{A_e^2}{T} ~~~.
\eeq
This is equivalent to assuming that the $A$ term
mediates a decay of one scalar into the other two.
It is not clear that this is kinematically allowed,
but I use this estimate anyway (I estimate the scattering
rate $\Gamma (\phi \phi \rightarrow \phi \gamma) \simeq 
10^{-3} g^2 A^2/T$, which gives a bound on $A$ that
is about an order of magnitude weaker.).  In the basis where
$m^2$ is diagonal, I need
\beq
\Gamma_A \sim 10^{-2} \frac{(A_e^{ij})^2}{T} < H ~~~~~~(i \neq j)
\eeq
for whichever lepton flavour $i$
one wishes  to conserve. This gives $A^{ij} < 10^{-5}$ GeV. 
This is not
a particularily strict bound; recall that
I have absorbed the yukawa into $A$
(see equation (\ref{soft})). Similar bounds
apply to the $R$-violating $A$ terms of
equaltion (\ref{Rsoft}).

Finally I need  to calculate a bound
on $B-L$ violating  trilinear interactions.
There are no basis confusions for these
interactions, so to conserve, for instance,
 $B/3- L_e$, I need all the trilinears
that violate $L_e$ and $B$ to satisfy
\beq
10^{-2} \lambda^2 T < H \simeq \frac{25 T^2}{m_{pl}}
\label{CCC}
\eeq
This means that $\lambda^{''}_{pqr} < 10^{-7}$
for all $p,q,r$,  $\lambda'_{epq}< 10^{-7}$
for all $p,q$, and $\lambda_{ejk},\lambda_{jke} < 10^{-7}$
for  $j,k \neq e$.

\subsection{summary of bounds on $R$ violating interactions}

Suppose that I define the Higgs
such that the three $B_i$ are zero.
Then to conserve $B/3 - L_e$, I need
\beq
\epsilon_e < 10-100 ~~{\rm keV},~~~~~~m^2_{H_1 L_e} < ( 60-90~{\rm MeV})^2
\eeq
for $\mu^2 \simeq m^2 \simeq B \simeq 100$ GeV.

I need all the trilinears and $A$ terms that violate $B/3 - L_e$
to respectively  satisfy
\beq
\lambda< 10^{-7}~~,~~~~ A < 10-100 ~{\rm keV}
\eeq

I also need $L_e$ flavour violation mediated
by the $R$ conserving soft masses to be
out of equilibrium, which will be the case if
\beq
\frac{m^2_{ei}}{m^2_{ii}} \lappeq 5 \times 10^{-2}~~, ~~~~
A_e^{ei} \lappeq 10- 100 ~{\rm keV}
\eeq

\section{if the BAU was not there...}

 I wrote in the introduction that I would
assume that the BAU was present in the thermal
soup above the EPT. This is of course
one of the loopholes
in the constraints that I am discussing;
if the BAU is {\it not} there, then there are
no constraints on $B-L$ violation.  So in this
section I will relax this assumption.

First I would like to work out over what temperature range
 the BAU needs to be ``not there''. The lower
end of this range is straightforward to find.
The $B+L$ violating electroweak effects drop
out of equilibrium at or shortly after the
electroweak phase transition. Once they are gone,
the BAU is no longer at risk. So the baryon
asymmetry can safely be present in the plasma
at $T \ll T_c$.

It can also be present at high temperatures, before the
electron yukawa comes into equilibrium. 
The interaction rate associated with a dimensionless
couplings constant $\lambda$ scales
linearly with the temperature: $\Gamma \sim \lambda^2 T$.
Since $H \sim T^2/m_{pl}$, this means that yukawa interactions
are out of thermal equilibrium
at sufficiently high temperatures,
and come into equilibrium as the
temperature drops. Since the electron yukawa
is the smallest coupling constant in the SM,
it is the last to come into equilibrium at
$T_{e_R} \sim 100$ TeV. 
The $e_R$ carries hypercharge, and the
Universe ought to be hypercharge neutral,
so if there is an asymmetry in the $e_R$,
and no interaction in equilibrium that
can transfer this asymmetry to other
particle species,  then asymmetries must remain
in other particles to ensure that
$Y=0$.  It is easy to check
from the relevant equations of chemical equilibrium
that this means a baryon asymmetry will remain
in the plasma. At $T \gg T_{e_R}$,
there will be other yukawa interactions out of equilibrium,
and additional symmetries \cite{IQ} that
can protect the baryon asymmetry, but 
 the singlet $e_R$ is the last particle
species to come into chemical equilibrium,
so can protect the baryon asymmetry  for the longest \cite{CDEO3}.
Any  $B-L$ violating operator
not involving the $e_R$ can be in equiliibrium
above $T_{e_R}$, without being able to eat the
BAU. This was not fully realised in the earlier
papers discussing bounds on non-renormalisable
$B -L$ violating operators, where strong bounds
were set on $D>4$ operators
by requiring that they be out
of equilibrium up to $T_{reheat}$, or
the temperature when the BAU was generated, or
the temperature when the $B+L$ violation came
into equilibrium.

Note that this whole argument assumes that there
is an asymmetry stored in the $e_R$. This is
maybe not so easy to arrange---for instance
the decay of heavy singlet (``right-handed'')
majorana neutrinos generates an asymmetry in
the lepton doublets, but not the charged
singlets. However, for setting generic bouds,
one must allow for the possibility that
some baryogenesis mechanism does generate
an  $e_R$ asymmetry, which can then
protect the baryon asymmetry down to
$T_{e_R}$. 

To approximately determine $T_{e_R}$,
one can assume that the $e_R$ comes
into chemical equilibrium when the decay
of the Higgs into a left handed and
a right-handed lepton is of order $H$ \cite{CDEO3}:
\beq
\Gamma \simeq 10^{-2} h_e^2 T \simeq H
\eeq
which gives $T_{e_R} \simeq 100$ TeV.
See \cite{CKO} for a careful
determination of $T_{e_R}$, that
includes the various decay and
scattering interactions mediated by $h_e$,
and the thermal masses of the participating particles.

So the baryon asymmetry is really only
``at risk'' over three decades in temperature:
$ T_c < T< 100$ TeV. If the asymmetry is
present in the thermal soup during this
period, it needs to be an asymmetry in $B-L$,
and  interactions violating at least
one of the $B/3 - L_i$ need to
be out of equilibrium during this
period.  This gives the bounds of section 2.

Alternatively, the BAU can  be not in the soup between
$ T_c < T< 100$ TeV.  The most obvious way
to do this is to generate the baryon asymmetry
at the electroweak phase transition (see, {\it e.g}
\cite{BAUrev,RS} for reviews of electroweak baryogenesis).
But this is not the only possibility; if we
turn out to be living in a corner of SUSY parameter
space that does not allow electroweak baryogenesis,
there are models where the baryon asymmetry is generated
in the late decay of  some particle, after the
EPT \cite{latedecay}. 

Another possibility is to generate the
baryon asymmetry early, and then ``hide it''
between $T_{e_R}$ and $T_c$. One 
way to do this \cite{DH}, 
is  to add additional particles
that carry baryon number to the SM or the SSM, 
decouple  them from
the rest of the SM particles at $T > T_{e_R}$
and have them decay after the EPT, returning the
baryon asymmetry that they carry to
the plasma of SM particles.  Another
possibilityy is to store some hypercharge in a 
coherent hypercharge magnetic field for
$T_c < T <T_{e_R}$ \cite{JS}. If there
is a sufficiently large asymmetry in the $e_R$,
it can act as a source for the hypercharge magnetic
field before $T_{e_R}$, and this field may
survive until the EPT.

Another somewhat related loophole is
to take  the up-quark yukawa to be
considerably smaller than  is usually
assumed \cite{up}. If  $h_u \lappeq 10^{-7}$
($m_u \lappeq 20 $ keV !), then 
an asymmetry in the $u_R$s could
play the same role as the $e_R$
asymmetry does, but all the way
down to the EPT. If $h_u$ did not
come into equilibrium until after the
EPT, and if the $u_R$s carried a
primordial asymmetry, then  this asymmetry
will be present after the EPT,
irrespective of how much
$B-L$ violation is in equilibrium
above the EPT (providing of course
that the $B-L$ violating operators
do not  involve $u_R$).

\section{other loopholes}

In this section, I list  the remaining loopholes.

\subsection{lepton mass effects}

In models
where  $n_{B-L} = 0$,
lepton mass effects can
protect the baryon asymmetry  from
SM $B+L$ violation,  provided that
there are lepton flavour asymmetries \cite{KRS}.  
A small baryon asymmetry will remain
below \cite{KRS} and above \cite{DKO}
the EPT, proportional to lepton mass differences
and flavour asymmetries. Since I am here
discussing what happens above
the EPT, the numerical factors in equation
(\ref{masseffects})  are for this case, where the lepton
masses are ``thermal''. In this subsection,
I neglect most of the numerical factors.

The number density
for particles in thermal equilibrium is
\beq
n  = \frac{2 g}{\pi^2} \int \frac{p^2 dp}{e^{(E-\mu)/T} \pm 1}
\eeq
where the $\pm$ is for fermions or bosons,
and the number density  $\bar{n}$ for anti-particles
has the sign in front of the chemical potential
$\mu$ flipped. Expanding $n - \bar{n}$ in
$\mu/T$ and  $m^2/T^2$, assuming that
both are small, gives
equation (\ref{nbarn}). As previously
noted, to correctly determine
the $O(\mu m^2)$ term, one must
calculate the free energy, and take
the derivative with respect to $\mu$.
However, I am not here interested in the
exact coefficient; see \cite{KS,DKO}
for numerical factors. 

The reason that these $(\mu m^2)$ contributions
are interesting is that they can preserve
a baryon asymmetry in the presence
of  the  $B+L$ violating interactions,
when $B-L = 0$, and there are
flavour asymmetries amoung  leptons
(also when $B+L$ and $B$ violation
are in equilibrium, and there are
flavour asymmetries).
They do not protect the BAU
if there are interactions in equilibrium
that take all three lepton chemical potentials
to zero. 

So suppose that the $B+L$ violating
interactions are in equilibrium (equation \ref{B+L}), and
that $n_{B-L} = 0$:
\beq
12 q_L - 3\sum_i \ell_i \left[1 + O\left( \frac{m_i^2}{T^2} \right) \right]
- 3H = 0
\label{end}
\eeq
 This could be 
the case if the BAU was generated in the
out-of-equilibrium decay of heavy
particles from a $B-L$ conserving GUT. 
There are two types of constraint
imposed on the plasma: the sum of the
chemical potentials 
participating in each interaction
must be zero, and certain charge
densities ($Y, B-L$) must be zero. 
If one includes the $O(\mu m^2)$ terms 
in the charge densities, one finds that
there is a small BAU in equilibrium.
This is due to the  non-zero 
chemical potentials and not equal masses for the different
lepton generations.

From the condition of charge neutrality, 
equation (\ref{Qem}), 
I can solve for the Higgs chemical
potential in terms $q_L$ and  $\sum_i \ell_i$:
\beq
H \simeq \frac{3 q_L - \sum_i \ell_i [ 1 + O(m^2_i/T^2)]}{7}
\simeq \frac{ 12 q_L - \sum_i \ell_i \times O( m^2_i/T^2)]}{7}
\eeq
(This assumes SM particle content, and
I have dropped $H m_i^2$ terms).  Substituted
into (\ref{end}), and using (\ref{B+L}) one finds
\beq
n_B \propto q_L T^2 \sim \sum_i \ell_i m_i^2
\label{remain}
\eeq 

$\sum_i \ell_i m_i^2 \neq 0$,
if there are lepton flavour asymmetries
($n_{L_i - L_j} \neq 0$), so
the remaining baryon asymmetry (\ref{remain})
is proportional to lepton mass differences,
and flavour asymmetries, see equation
 (\ref{masseffects}). It is clear from 
 (\ref{masseffects}), where all the
numerical factors are included, that
one needs very large lepton flavour asymmetries
($\sim 10^{-2}- 10^{-3}$) to preserve
a sufficiently large baryon asymmetry.
An interesting possibility suggested
in \cite{DR} is to use the slepton mass
differences in a SUSY model,  because
these could be much larger.

\subsection{modified cosmologies}

In this subsection I  briefly review
some modifications of cosmology that
allow one to escape the bounds discussed
in section 2. 

One possibility is low $T_{reheat}$ models;
if the Universe after inflation
reheats to a temperature $T_{reheat} < T_c$,
then one never has to worry about having
$B+L$ and $B-L$ violating operators
simultaneously present. However, one does
have to worry about how to generate
the baryon asymmetry in such models.
This is not neccesarily a problem;
one can, for instance, generate an asymmetry from
the decay of the inflaton, or in some
cases via the Affleck-Dine mechanism \cite{A+D}.
In fact, the reheat temperature after
Affleck-Dine baryogensis is often low.

Another possibility  is to modify the expansion rate
of the Universe. If the Universe expanded
somewhat faster, the $e_R$ could
protect the BAU all the way down to $T_c$.
This can be accomplished by, for instance, adding 
a component to the energy density of the
Universe that scales as $1/a^6$ \cite{JP}.
(This is not as
far-fetched as it might sound;
a scalar field in the right shaped potential
can behave this way.)

It is also possible to ``get rid of'' the
$B+L$ violating electroweak interactions. For
instance, if there is a SU(2) doublet with
a large vev above the EPT, it would
give a mass to the $W$, so the
rate for $B+L$ violation would be 
exponentially suppressed. This
could occur in an
an Affleck-Dine baryogenesis scenario,
when the condensate carries a
large baryon asymmetry \cite{DMO},
and therefore does not decay
until $T \sim 100$ GeV. 

\section{summary}

The argument leading to cosmological
bounds on $B-L$ violating interactions
goes as follows. Assume that the
baryon asymmetry was created before the
electroweak phase transition as an asymmetry
in at least one of the three $B/3-L_i$, and that
it is present in the thermal soup above
the electroweak phase transition.   The
electroweak $B+L$ violating interactions
are  operating in this plasma.
If there are also interactions
in equilibrium that violate all of the
$\{B/3 - L_i\}$, the baryon
asymmetry will be washed out. 
Since it is supposed to be around
today, this is undesirable. Therefore
one requires that interactions
violating one of the $B/3 - L_i$
be out of thermal equilibrium, so they
cannot eat the baryon asymmetry. 

This generically gives bounds of
order $\lambda < 10^{-7}$ on
``yukawa-type'' coupling constants,
and $m \lappeq $ keV-100 MeV for
different masses. The mass
bounds can be slightly tricky
to calculate. These constraints
apply to interactions that violate
$B$ and one particular lepton
flavour.

There are various loopholes to this
argument. One can
avoid them completely by
not having the BAU present in the
plasma for 100 TeV $ > T > $ 100 GeV.
This is approximately when it is at risk from
the $B+L$ and $B-L$ violating interactions.
The most obvious way to not have the
 baryon asymmetry present 
 is to generate it
at the electroweak phase transition.
There are other models where
the BAU is generated after the
phase transition, or where it is
generated at very high temperatures
and hidden for 100 TeV $ > T > $ 100 GeV.

It is also possible to modify the cosmological
model to avoid these bounds. If the reheat
temperature after inflation is below
that of the electroweak phase transition,
there is no opportunity to wash out the
BAU. (One has fewer ways of making it
though). If there electroweak
$B+L$ violating
interactions are turned off above
the phase transition, then an asymmetry in $B+L$
can survive. If the expansion rate of the
Universe is increased, the slowest
Standard Model interaction, which is
the electron yukawa, might not come
into equilibrium until after the
phase transition. If there is
an asymmetry in the $e_R$, this means
that an asymmetry must remain in the
baryons, because the Universe should be
hypercharge neutral.

It is possible to avoid the strong bounds
on $B$ violating operators, or the 
conclusion that the asymmetry must 
carry $B-L$, via lepton mass
effects. If there are large lepton
flavour asymmetries, and lepton flavours
are conserved, then a small baryon asymmetry
proportional to $ m_{\tau}^2/T^2 \times n_{L_{\tau} - L_i}$
will remain, even if $n_{B-L} = 0$
or if there are $B$ violating interactions
in equilibrium.


\begin{thebibliography}{222222}
\bibitem{causality} see, {\it e.g.}, (and references therein)
   A.G. Cohen, A. De Rujula, S.L. Glashow,
  {\it  Astrophys.J.} {\bf 495} (1998) 539. 
   astro-ph/9707087


\bibitem{BAUrev} see {e.g.} \\
  A .D. Dolgov, 
  {\it  Phys.Rept.} {\bf 222} (1992) 309. \\
    A. Riotto, 
  talk at {\it ICTP Summer School in High-Energy 
  Physics and Cosmology},  Trieste, Italy,  July 1998. 
  hep-ph/9807454. 

\bibitem{RS}
  V.A. Rubakov, M.E. Shaposhnikov, 
  {\it  Usp.Fiz.Nauk} {\bf 166} (1996) 493-537. ( Phys.Usp.39 (1996) 461. 
  hep-ph/9603208 

\bibitem{KT} E.R. Kolb, M. Turner {\it `` The Early Universe''},
Addison-Wesley, 1990.

\bibitem{Sakharov}
  A.D. Sakharov. 1967. 
  {\it  Pisma Zh.Eksp.Teor.Fiz.} {\bf 5} (1967) 32-35.
  ( JETP Lett.5 (1967) 24.

\bibitem{Ross}  for a discussion of B and L violating
       operators on the Standard Model and in supersymmetry, 
       see, $e.g.$ G.G.Ross, {\it Grand Unified Theories},
       Benjamin-Cummings, Menlo Park, CA,
       1985, chapters 7 and 11.

\bibitem{'tHooft}
G. 't Hooft, {\it Phys. Rev. Lett. } { \bf 37} (1976) 8;
G. 't Hooft, {\it Phys. Rev.} { \bf D14} (1976) 3432.


\bibitem{KRSa}
  V.A. Kuzmin, V.A. Rubakov, M.E. Shaposhnikov,  
   {\it Phys.Lett.} {\bf 155B} (1985) 36.

\bibitem{Arnold}
P. Arnold, D. Son, L. Yaffe, {\it Phys. Rev.} {\bf D 55} (1997) 6264.

\bibitem{EPT} 
K. Kajantie, M. Laine, K. Rummukainen, M. Shaposhnikov,
{\it Nucl.Phys.} {\bf B466} (1996) 189. 
 hep-lat/9510020 



\bibitem{GUTBAUrev} see, {\it e.g.}
  N. Fry, K. A. Olive, M. S. Turner,
  {\it Phys.Rev.} {\bf D22} (1980) 2977,
  {\it Phys.Rev.} {\bf D22} (1980) 2953. \\
   E. W. Kolb, S. Wolfram, 
  {\it Nucl.Phys.} {\bf B172} (1980) 224,  Erratum-{\it ibid.} 
  {\bf B195} (1982)  542.


\bibitem{FYBAU}
   M. Fukugita, T. Yanagida, 
   {\it  Phys.Lett.} {\bf 174B} (1986) 45. 


\bibitem{FY}
  M. Fukugita, T. Yanagida, 
  {\it  Phys.Rev.} {\bf D42} (1990) 1285.

 
\bibitem{NB}
   A. E. Nelson, S. M. Barr, 
  {\it  Phys.Lett.} {\bf B246} (1990) 141.


\bibitem{FGLP}
   W. Fischler, G.F. Giudice, R.G. Leigh, S. Paban, 
   {\it  Phys.Lett.} {\bf B258} (1991) 45. 


\bibitem{CDEO1}B. A. Campbell, S. Davidson, J. Ellis, K.A. Olive
  {\it Phys.Lett.} {\bf B256} (1991) 457.


\bibitem{CDEO2}B.A. Campbell, S. Davidson, J. Ellis, K. A. Olive 
  {\it Astropart.Phys.} {\bf 1}  (1992) 77. 

\bibitem{DR}
   H. Dreiner, G.G. Ross, 
   {\it  Nucl.Phys.} {\bf B410} (1993) 188. 

\bibitem{CDEO3}B. A. Campbell, S. Davidson, J. Ellis, 
  K.A. Olive,
  {\it Phys.Lett.} {\bf B297} (1992) 118.

\bibitem{IIMS} 
  T. Inui, T. Ichihara, Y. Mimura, N. Sakai
  {\it Phys.Lett.} {\bf B325} (1994) 392. hep-ph/9310268 


\bibitem{other}
G. Gelmini, T. Yanagida, {\it Phys. Lett.} {\bf B 294} (1992) 53.\\
W.Buchmuller, T. Yanagida,
{\it  Phys.Lett.} {\bf B302} (1993) 240.\\
J.M. Cline, K. Kainulainen, K. A. Olive,
{\it  Astropart.Phys.} {\bf 1} (1993) 387. 
 hep-ph/9304229 \\
C.E. Vayonakis,
{\it Phys.Lett.} {\bf B286} (1992) 92.\\
A. Antaramian, L.J. Hall, A. Rasin,
{\it  Phys.Rev.} {\bf D49} (1994) 3881.
 hep-ph/9311279 \\
A. Ganguly, J.C. Parikh, U. Sarkar,
{\it Phys.Lett.} {\bf B385} (1996) 175. 
 hep-ph/9408271 

\bibitem{herbi} H. Dreiner, to be published in 
   '{\it Perspectives on Supersymmetry} ', Ed. by G.L. Kane, 
    World Scientific;  hep-ph/9707435. 


\bibitem{IR}
  L.E. Ibanez, G. G. Ross,
  {\it Nucl.Phys.} {\bf B368} (1992) 3. 


\bibitem{CKO}
  J.M. Cline, K Kainulainen, K. A. Olive, 
  {\it Phys.Rev.Lett.} {\bf 71} (1993) 2372. 
   hep-ph/9304321; 
  {\it Phys.Rev.} {\bf D49}  (1994) 6394. 
   hep-ph/9401208 


\bibitem {HT}
J.A. Harvey, M. S. Turner
{\it  Phys.Rev.} {\bf D42} (1990) 3344.



\bibitem{KS}
   S.Yu. Khlebnikov, M.E. Shaposhnikov, 
  {\it  Phys.Lett.} {\bf B387} (1996) 817.  \\
 S.Yu. Khlebnikov, M.E. Shaposhnikov,
 {\it Nucl.Phys.} {\bf B308} (1988) 885. 

 
\bibitem{KRS}
   V.A. Kuzmin, V.A. Rubakov, M.E. Shaposhnikov,  
   {\it Phys.Lett.} {\bf 191B} (1987) 171.



\bibitem{DKO}
   S. Davidson, K. Kainulainen, K.A. Olive, 
   {\it  Phys.Lett.} {\bf B335} (1994) 339. 
    hep-ph/9405215

\bibitem{DE1}
  S. Davidson, J. Ellis, 
  {\it Phys.Lett.} {\bf B390} (1997) 210-220. 
  hep-ph/9609451 

\bibitem{DE2}
   S. Davidson, J. Ellis, 
  {\it  Phys.Rev.} {\bf D56}  (1997) 4182. 
   hep-ph/9702247 

\bibitem{D}
  S. Davidson, to be published in {\it Phys. Lett.} {\bf B}.
  CERN-TH-98-161.

\bibitem{IQ}
  L. E. Ibanez, F. Quevedo, 
 {\it  Phys.Lett.} {\bf B283} (1992) 261. 
  hep-ph/9204205 

\bibitem{latedecay}
S.Dimopoulos, L.J. Hall, {\it Phys. Lett.} {\bf B 196} (1987) 135; 
  J. Cline, S. Raby, {\it Phys. Rev.} {\bf D 43} 
  (1991) 1781; 
  R. J. Scherrer, J. Cline, S. Raby, D. Seckel,
  {\it Phys.Rev.} {\bf D44} (1991) 3760;
  A. Masiero, A. Riotto, {\it Phys.Lett.} {\bf B289} (1992) 73.  

\bibitem{DH}
  S. Davidson, R. Hempfling,
  {\it Phys.Lett.} {\bf B391} (1997) 287. 
   hep-ph/9609497 


\bibitem{JS} 
  M. Joyce, M. Shaposhnikov,
  {\it Phys.Rev.Lett.} {\bf 79} (1997) 1193.
  astro-ph/9703005 
 
\bibitem{up} for a discussion, see for instance,\\
J. E. Kim,
{\it Phys.Rept.} {\bf 150} (1987) 1. \\
T. Banks, Y. Nir, N. Seiberg,
{\it Gainesville 1994, Yukawa couplings and the origin of mass},
 Gainesville, FL,  Feb 1994. 
 hep-ph/9403203 



\bibitem{A+D} see, for instance, 
  I. Affleck, M.Dine, {Nucl. Phys.} {\bf B 249}
  (1985) 361 ; 
  J. Ellis , K. Enqvist, D.V. Nanopoulos, K.A. Olive, 
  {\it Phys. Lett.} {\bf  B191} (1987) 343;
  M.Dine, L. Randall, S. Thomas, {\it Phys. Rev. Lett.}
  {\bf 75} (1995) 398;
  {\it ibid.} {\it Nucl. Phys.} {\bf B458} (1996) 291;
  M. Gaillard, H. Murayama, K.A. Olive, {\it Phys.Lett.} {\bf B355}
  (1995) 71.

\bibitem{JP}
  M. Joyce, T. Prokopec, 
 {\it  Phys.Rev.} {\bf D57} (1998) 6022. 
 hep-ph/9709320 

\bibitem{DMO}
   S. Davidson, H. Murayama, K. A. Olive,
  {\it  Phys.Lett.} {\bf B328} (1994) 354. 
   hep-ph/9403259 
\end{thebibliography}
\end{document}